# The Synthetic Hilbert Space of Laser-Driven Free-Electrons

**Guy Braiman†, Ori Reinhardt†, Chen Mechel, Omer Levi, and Ido Kaminer**

Department of Electrical and Computer Engineering and Solid State Institute, Technion - Israel Institute of Technology, 32000 Haifa, Israel

† Equal contribution

**Recent advances in laser interactions with coherent free electrons have enabled to shape the electron's quantum state. Each electron becomes a superposition of energy levels on an infinite quantized ladder, shown to contain up to thousands of energy levels. We propose to utilize the quantum nature of such laser-driven free electrons as a "synthetic Hilbert space" in which we construct and control qudits (quantum digits). The question that motivates our work is what qudit states can be accessed using electron–laser interactions, and whether it is possible to implement any arbitrary quantum gate. We find how to encode and manipulate free-electron qudit states, focusing on dimensions which are powers of 2, where the qudit represents multiple qubits implemented on the same single electron – algebraically separated, but physically joined. As an example, we prove the possibility to fully control a 4-dimenisonal qudit, and reveal the steps required for full control over any arbitrary dimension. Our work enriches the range of applications of free electrons in microscopy and spectroscopy, offering a new platform for continuous-variable quantum information.**

## Introduction

Physical systems for quantum information processing[1] have been widely explored in the last decades[2-6], motivated by efficient quantum algorithms such as Shor's algorithm[7]. A variety of platforms are being investigated for implementing Hamiltonians that enable storing and manipulating a high-dimensional quantum state. The vast majority of platforms that can encode quantum information use qubits, 2-level systems that each encode a "unit" of quantum information. Most of these platforms rely on encoding quantum information on the Hilbert space of bound electrons, such as ion traps[2], artificial atoms implemented on superconducting circuits[5], Rydberg atoms[8], and electron-spin systems[6]. A recent proposal[9] that was demonstrated experimentally[10] suggested going beyond *bound electrons* as carriers of quantum information, and instead encode qubits on *free electrons* by using their interaction with ultrafast laser pulses.

Quantum information can be contained within *qudits* (quantum digits) rather than qubits. A qudit is a $d$-dimensional quantum system, with a Hilbert space that is spanned by the basis $|0\rangle, |1\rangle, \dots, |d-1\rangle$. One of the ways to physically implement qudits is through a mathematical projection from a larger finite/infinite Hilbert space that is more convenient to manipulate. We hereafter refer to these projected spaces as "synthetic Hilbert spaces", as they do not appear naturally in experiments but nonetheless can be used for quantum computation. We use this term of a synthetic Hilbert space to highlight this induced algebra that is added here beyond the added dimension. This way, the electron energy degree of freedom provides both (1) an additional dimension (i.e., a synthetic dimension), and (2) an induced qudit algebra (e.g., as in a high dimensional spin). Ideas of implementing qudit algebra in such synthetic Hilbert spaces were investigated within molecular energy levels[11,12], in the time-energy domain of a single photon[13-15], or in its orbital angular momentum (OAM)[16]. Other notable realizations encode the quantum information in the path photons take, for example, in free space optical setups or in silicon waveguides[4,17-20]. The limit of $d \to \infty$ merges with the area of continuous-variable quantum information[21], which can encode a qubit on the states of a harmonic oscillator in a fault tolerant manner[22,23]. The generation of such fault-tolerant states is now at the forefront of research toward photonic quantum computation[24].

Our work proposes a novel platform for encoding continuous-variable quantum information – on the synthetic Hilbert space created from the discrete energy levels of laser-driven free electrons. We show how a laser-driven free electron can be imprinted by a *qudit* of arbitrary finite size, and how its state can be manipulated. This form of information encoding in free electrons is significantly different from the conventional forms seen in bound-electron and photonic systems not only in the physics, but also in the mathematics – as it is a type of continuous-variable system that is not described by a harmonic oscillator. We investigate the mathematical features of this synthetic Hilbert space and discuss the prospects of its physical implementation.

In this synthetic Hilbert space, we supply a proof of the ability for *full control* of a qudit of size 2 or 4 and develop methods toward the full control of qudits of arbitrary dimensions on the same single electron. Our work promotes new capabilities based on transaction of quantum information at the interface between electron microscopy and quantum information, toward free-electron-based quantum communication and quantum electron microscopy[25-28]. Looking forward, the ability to encode high-dimensional quantum states on a single electron suggests the possibility to implement new schemes of fault-tolerant quantum information on such platforms.

Our proposal to use free electrons as quantum information carriers has become feasible due to experimental advances in ultrafast electron microscopy (UTEM)[29]. Specifically, the technique of photon-induced nearfield electron microscopy (PINEM)[30] demonstrated the feasibility of the necessary coherent control of the electron energy spectrum in our quantum information schemes (figure 1a). Our work relies on an approach for free-electron coherent control based on multiple electron-laser interactions, separated by free space propagation to accumulate phases. This technique can be applied with a single laser harmonic[31,32] as well as using several harmonics[33,34] for improved control of the spectrum. The PINEM technique has spawned a plethora of novel directions of theory and experiment in recent years[9,10,33,35-37]. All of these rely fundamentally on the electron undergoing a quantum walk on an unbounded and bi-directional energy ladder, with uniform spacing that equals the photon energy $\hbar\omega$[30,38,39] (figure 1b).

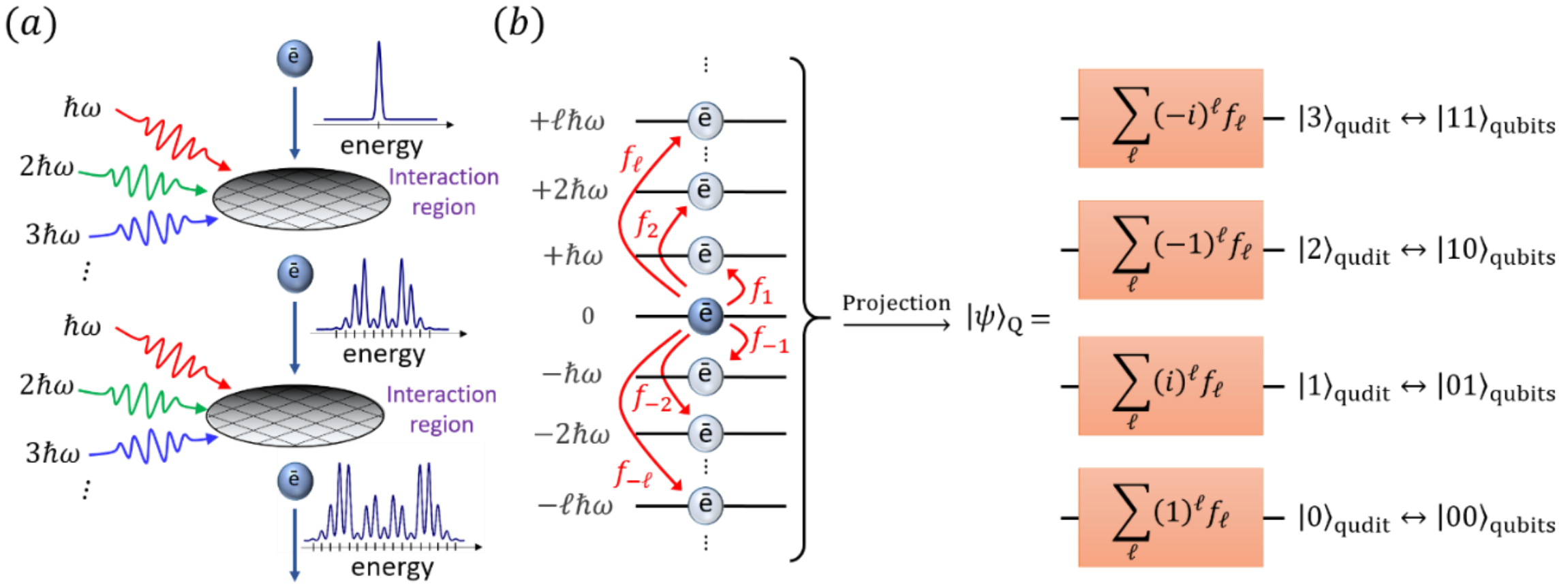

**Figure 1. Illustration of the synthetic Hilbert space of a laser-driven free electron. (a)** Interactions between a free electron and an arbitrary laser pulse comprised of harmonics of the fundemental frequency $\omega$. The electron undergoes sequential interactions with the laser in different interaction regions, and as a result its energy spectrum changes into coherent superposition states. **(b)** The quantizied energy ladder, relative to the electron's initial

energy. During the interaction, the electron performs a quantum walk on the ladder and ends up in a superposition of energy levels $\ell\hbar\omega$ with amplitudes $f_\ell$. We use this synthetic dimension to encode the qudit state $|\psi\rangle_{\mathrm{Q}}$, where the weight of each basis vector $|i\rangle_{\mathrm{Q}}$ of the qudit is determined according to the sum of all relevent probability amplitudes with the appropiate coeffcient. Upon that, we can define 2-qubit states and other qudit states on a single electron.

## Results: the general framework

We translate the electron's Hilbert space into a well-defined *qudit of finite size*. This translation effectively turns the electron physical energy space into a *synthetic* dimension of a specific desired size. Our recent work[9] proved the feasibility of realizing a single *qubit* on a free electron. We algebraically translated the quantized ladder of electron energy states into a 2-level system, and we proved that this creates a completely valid qubit – a 2-level system that is fully controllable. Our present work proposes a realization of a single qudit of arbitrary finite size on a single free electron. This qudit can be used as a Hilbert space for 1-qudit or *$n$-qubit computation on a single free electron*, by choosing the qudit dimension to be $2^n$.

We next present our implementation scheme and discuss its interesting properties. A primary ingredient in our realization is the PINEM interaction, whereby an energetic electron interacts with a laser pulse composed of harmonics of a fundamental frequency $\omega$[33,36]. The electron has a narrow energy spread $\Delta E_0$, such that the electron's initial energy $E_0$ satisfies $E_0 \gg \hbar\omega > \Delta E_0$. We denote by $\ell$ the number of photons gained/lost by the electron, so we can write the electron state in the discrete, infinite, orthonormal energy basis; $|\psi\rangle = \sum_\ell \psi_\ell |\ell\rangle$. The PINEM interaction operator is unitary in this basis. For a PINEM interaction consisting of two frequencies $\omega_1 = \omega$ and $\omega_2 = 2\omega$, its matrix representation following Refs. [9,40] and [33,35,36] is

$$U_{PINEM}(g_1, g_2) = \exp\begin{pmatrix} \ddots & g_1^* & g_2^* & & \\ -g_1 & 0 & g_1^* & g_2^* & \\ -g_2 & -g_1 & 0 & g_1^* & g_2^* \\ & -g_2 & -g_1 & 0 & g_1^* \\ & & -g_2 & -g_1 & \ddots \end{pmatrix}. \quad (1)$$

The parameters $g_{1,2}$ are the coupling constants/PINEM fields[37,40] corresponding to the harmonics $\omega_{1,2}$. The PINEM coupling constant is a dimensionless complex parameter, whose magnitude indicates the interaction strength between the electron and the laser's electric field[9,41,42].

The two-frequencies PINEM allows us to fully realize a 4-dimensional qudit, as proved below. Applying the PINEM interaction operator given in equation (1) on an electron energy component $|\ell'\rangle$ yields $U_{PINEM}|\ell'\rangle = \sum_\ell f_\ell |\ell' + \ell\rangle$, with probability amplitudes $f_\ell$ that depend on the coupling constants and can be calculated analytically (see supplementary material for details). An important result is that different PINEM operators commute, regardless of the interaction strengths or driving laser frequencies[33,36]. Since the qubits/qudit algebra is non-commutative, this implies the necessity of using another operator to implement the universal set of quantum operations and reach every quantum state in Hilbert space.

The necessary ingredient can be achieved by free-space propagation (FSP) of the electron between laser interactions. This propagation yields different phase accumulation by each energy component $|\ell\rangle$, as was demonstrated inside electron microscopes and other systems[32-34]. The phase accumulation of each energy component along a propagation distance $z$ is given by $\phi_\ell = \phi\ell^2 = 2\pi\frac{z}{z_D}\ell^2$, with the characteristic length $z_D = 4\pi\beta^2\gamma^3\frac{\omega_C}{\omega}\frac{v}{\omega}$, where $\beta = v/c$ is the electron normalized velocity, $\gamma$ the electron Lorentz factor, and $\omega_\mathrm{C}$ the Compton frequency, given by $\omega_\mathrm{C} = mc^2/\hbar \approx 7.8 \times 10^{20}[\mathrm{rad/s}]$[9]. We can represent the FSP operator as an infinite, unitary matrix in the following way:

$$U_{FSP}(\phi) = \exp\begin{pmatrix} \ddots & & & & & & \\ & e^{-4i\phi} & & & & & \\ & & e^{-i\phi} & & & & \\ & & & 1 & & & \\ & & & & e^{-i\phi} & & \\ & & & & & e^{-4i\phi} & \\ & & & & & & \ddots \end{pmatrix}. \quad (2)$$

Crucially, since FSP operates in a different way on different energy states, it does not commute with the PINEM operator, which has energy-translational invariance. This is the key for the realization of a universal set of quantum gates using combinations of PINEM and FSP. Interestingly, this concept of "successive separated operations" appears in several other platforms of qubit information processing, such as coherent control of nuclear magnetic resonance[43]. Many of these platforms use combinations of electromagnetic pulses, mostly radiofrequency and microwaves, separated by appropriate time delays, analogously to PINEM and FSP.

We now address the challenges of realizing a 4-dimensioanl qudit. We extend the approach first proposed in [9] and modify it to help prove gate universality. The 4-dimensional qudit state can be defined using $|\psi\rangle = \sum_\ell \psi_\ell |\ell\rangle$ in the following way:

$$
|\psi\rangle_{\mathrm{Q}} \equiv \begin{pmatrix} \alpha_0 \\ \alpha_1 \\ \alpha_2 \\ \alpha_3 \end{pmatrix} = \frac{1}{2} \begin{pmatrix} \sum_{\ell} (1)^{\ell} \psi_{\ell} \\ \sum_{\ell} (-i)^{\ell} \psi_{\ell} \\ \sum_{\ell} (-1)^{\ell} \psi_{\ell} \\ \sum_{\ell} (i)^{\ell} \psi_{\ell} \end{pmatrix}, \tag{3}
$$

using the basis for the qudit's Hilbert space $|0\rangle_{\mathrm{Q}} = (1 \quad 0 \quad 0 \quad 0)^T$, $|1\rangle_{\mathrm{Q}} = (0 \quad 1 \quad 0 \quad 0)^T$, $|2\rangle_{\mathrm{Q}} = (0 \quad 0 \quad 1 \quad 0)^T$ and $|3\rangle_{\mathrm{Q}} = (0 \quad 0 \quad 0 \quad 1)^T$.

Equation (3) describes a projection from the electron's infinite space into the qudit's finite subspace. The projection uses all the energy components and thus increases robustness for noise in both gate operations and measurements. This infinite-to-finite projection could be viewed as a type of error-correction method. We note that a normalized electron state describing a finite-energy physical state results in a normalized qudit state. The same Hilbert space also represent 2 qubits, for example by translating $|0\rangle_{\mathrm{qudit}} \leftrightarrow |00\rangle_{\mathrm{qubits}}$, $|1\rangle_{\mathrm{qudit}} \leftrightarrow |01\rangle_{\mathrm{qubits}}$, $|2\rangle_{\mathrm{qudit}} \leftrightarrow |10\rangle_{\mathrm{qubits}}$ and $|3\rangle_{\mathrm{qudit}} \leftrightarrow |11\rangle_{\mathrm{qubits}}$.

We proceed to construct a universal set of 2-qubit gates using the PINEM and FSP operators that were presented in equations (1) and (2). We denote a 1-qubit gate $U$ operating on the $i$-th qubit ($i = 1,2$) by $U_i$. For example, the Hadamard gate $H$ operating on the first qubit is described by the operator $H_1 = H \otimes \mathbb{I}_2$ and on the second qubit by $H_2 = \mathbb{I}_2 \otimes H$, using the unity operator $\mathbb{I}_2$. To proceed, we analyze the algebraic form of the physical operators. One can prove that the qudit space is closed under the PINEM interaction given in equation (1) (see supplementary material section 4.1.6). For the FSP operator, it is not true in the general case; the FSP operator is unitary in the qudit space only for distances $n\frac{z_D}{8}$ for integer $n$ and $U_{FSP}\left(n\frac{z_D}{8}\right) = \left(U_{FSP}\left(\frac{z_D}{8}\right)\right)^n$. Therefore, every relevant FSP operation can always be written as a power of $U_{FSP}\left(\frac{z_D}{8}\right)$.

To prove the universality for this set of gates, we can express the PINEM operator and FSP operator in the 4-dimensional qudit space (see supplementary material for details). We obtain universality by using two PINEM frequency harmonics $\omega_1 = \omega$ and $\omega_2 = 2\omega_2$, corresponding to coupling constants $g_1$ and $g_2$, respectively. The effects of higher harmonics are included in the effects of $\omega_{1,2}$. Following our choice of the state definition in equation (3), these two harmonics define three degrees of freedom, which are expressed within the real and imaginary parts of $g_1$ and the imaginary part of $g_2$. To simplify the algebra, we define the parameters $\theta \equiv 2\mathrm{Im}\{g_1\}, \phi = 2\mathrm{Re}\{g_1\}$ and $\gamma = \mathrm{Im}\{g_2\}$, i.e. $g_1 \equiv \frac{\theta}{2} + i\frac{\phi}{2}$ and $g_2 \equiv \mathrm{Re}\{g_2\} + i\gamma$. Writing the gates in these notations:

$$U_{PINEM}(\phi,\theta,\gamma) = \begin{pmatrix} e^{-i(\gamma+\phi)} & & & \\ & e^{i(\gamma-\phi)} & & \\ & & e^{i(\theta-\gamma)} & \\ & & & e^{i(\phi+\gamma)} \end{pmatrix}, \tag{4a}$$

$$U_{FSP}\left(z = \frac{z_D}{8}\right) = \frac{1}{2}\begin{pmatrix} e^{-\frac{i\pi}{4}} & 1 & e^{i\frac{3\pi}{4}} & 1 \\ 1 & e^{-\frac{i\pi}{4}} & 1 & e^{i\frac{3\pi}{4}} \\ e^{i\frac{3\pi}{4}} & 1 & e^{-\frac{i\pi}{4}} & 1 \\ 1 & e^{i\frac{3\pi}{4}} & 1 & e^{-\frac{i\pi}{4}} \end{pmatrix}. \tag{4b}$$

This set of matrices is sufficient for generating the 2-qubit algebra, as we explicitly show in the supplementary material. The proof starts by generating the 1-qubit Hadamard gate $H_1$, which enables the implementation of the SWAP and $\mathrm{CNOT}_{2\to1}$ gates. Using $H_1$ and SWAP, one can generate both the second Hadamard gate, $H_2$ and the complementary CNOT gate, $\mathrm{CNOT}_{1\to2}$. The proof continues by generating the 1-qubit rotation matrices about the $z$-axis, $R_{z,1}$ and $R_{z,2}$, using a single PINEM gate, and realizing the rotation gates about the y-axis, $R_{y,1}$ and $R_{y,2}$.

Having the above matrices as building blocks of a universal set, one can rely on the work of Vatan and Williams[44] to construct an arbitrary 4-dimensional gate. Their work provides a method for constructing an *optimal* quantum circuit for any 2-qubit gate (these constructions are optimal with respect to the number of the CNOT gates and the number of $y$- and $z$- single-qubit rotations). They show that any gate requires at most 3 CNOT gates and 15 elementary one-qubit gates.

## Results: implementations of quantum gates

To further demonstrate the concept, we present the full designs of two types of computation: (1) achieving arbitrary-single qubit rotation, and (2) entangling 2-qubit to a Bell state. For an example computation of type (1), figure 2a presents a design of the gates $R_{z,1}\left(\frac{\pi}{2}\right)R_{z,2}\left(-\frac{\pi}{2}\right)$, followed by two consecutive $R_{x,1}\left(\frac{\pi}{4}\right)$ gates. The first gates can be implemented using PINEM, by an appropriate choice of $g_1 = \frac{\pi}{8}(1+i), g_2 = \frac{15}{16}\pi i$. Figure 2c presents the implementation of $R_{x,1}\left(\frac{\pi}{4}\right)$, which needs both PINEM and FSP, using the identity $HTH = R_x\left(\frac{\pi}{4}\right)$. Figure 2b displays the Bloch sphere representation of this quantum computation. The initial electron state is the mono-energetic electron state ($\psi_\ell = \delta_{\ell,0}$) that we denote by $|0\rangle$. This state is also called the zero-loss peak that can be quite monoenergetic in practical electron microscopes, having a narrow energy spread $\Delta E < \hbar\omega$. According to equations (3) and (4), the projection of this energy state onto the qubits subspaces is given by $|\psi\rangle_{\mathrm{Q}} = \frac{|0\rangle_{\mathrm{Q}_1}+|1\rangle_{\mathrm{Q}_1}}{\sqrt{2}} \otimes \frac{|0\rangle_{\mathrm{Q}_2}+|1\rangle_{\mathrm{Q}_2}}{\sqrt{2}}$. The computation results in the state $|0\rangle_{\mathrm{Q}_1} \otimes \frac{|0\rangle_{\mathrm{Q}_2}-i|1\rangle_{\mathrm{Q}_2}}{\sqrt{2}}$.

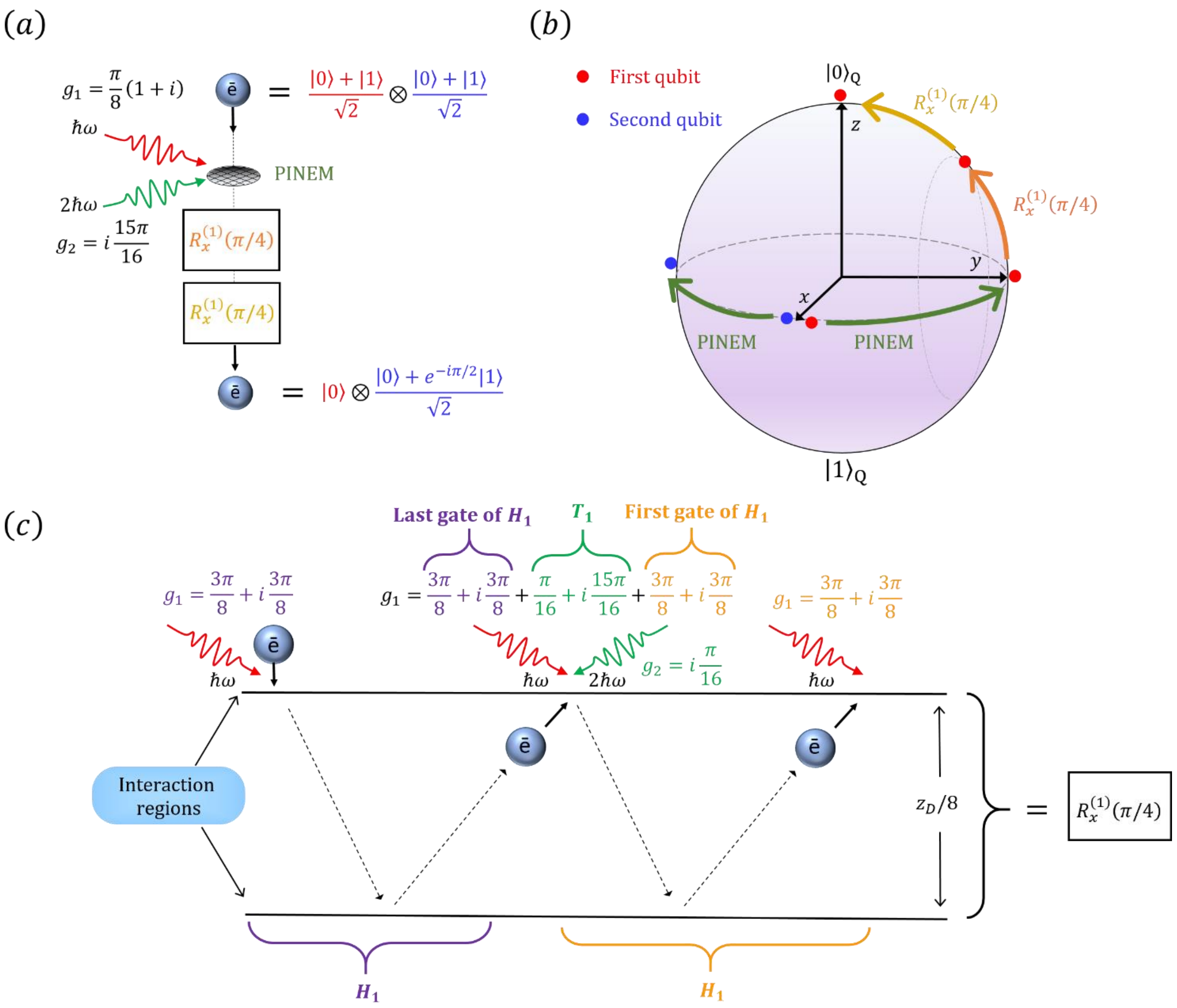

**Figure 2: Example of two-qubit gates on a single electron**, emphasizing the separation of the qudit space to independent qubit subspaces. The proposed gate performs independent operations on different qubit subspaces of the qudit. **(a)** Implementation of two 1-qubit quantum gates on two independent subspaces on the same free electron. The elecron undergoes a PINEM interaction, which translates to tensor product of two 1-qubit rotation matrices of $\pi/2$ about the $\hat{z}$-axis in the qubits space. Then, the gate $R_{x,1}(\pi/4)$ is applied twice in a row, changing the electron qubits state from $\frac{|0\rangle+|1\rangle}{2} \otimes \frac{|0\rangle+|1\rangle}{2}$ to $|0\rangle \otimes \frac{|0\rangle-i|1\rangle}{2}$. **(b)** Geometrical representation of the free-electron state, overlaying both qubit subspaces on the same Bloch sphere. The trajectories mark the procedure shown in (b). The state in the first qubit subspace is marked in red dots and the state in the second qubit subspace is marked in blue dots. Both qubit subspaces are rotated around the $\hat{z}$-axis due to the PINEM interaction. Then, applying the gate $R_{x,1}(\pi/4)$ twice in a row shifts the state of the second qubit subspace to the north pole $|0\rangle_{Q_2}$. Overall, the operation on the qudit is equivalent to $R_y(-\pi/2)$ on the first qubit subspace and $R_z(-\pi/2)$ on the second qubit subspace. **(c)** Construction of the $R_{x,2}(\pi/4)$ gate, which operates on one qubit subspace without altering the rest of the electron state (rotation by $\pi/4$ about the $\hat{x}$-axis, counter-clockwise).

For an example computation of type (2), figure 3 presents the construction of the maximally entangled Bell-state $|\tilde{\psi}_f\rangle_Q = |\beta_{00}\rangle_Q \equiv \frac{(|00\rangle_Q + |11\rangle_Q)}{\sqrt{2}}$, starting from the state $|\tilde{\psi}_i\rangle_Q = |00\rangle_Q$. This could be accomplished by applying the following computation:

$$|\tilde{\psi}_i\rangle_Q = |00\rangle_Q \xrightarrow{H_2} \frac{1}{\sqrt{2}}\left(|00\rangle_Q + |01\rangle_Q\right) \xrightarrow{\mathrm{CNOT}_{2\to1}} \frac{1}{\sqrt{2}}\left(|00\rangle_Q + |11\rangle_Q\right) \equiv |\beta_{00}\rangle_Q.$$

For completeness, the aspect of the qudit state initialization should also be considered. The mono-energetic state $|0\rangle$ can be produced in a UTEM by electron photo-emission with a low enough energy spread (e.g., [29,45,46]). When the state is written in the PINEM-basis, creating the state $|0\rangle_Q$ requires to initialize the electron state such that $\alpha_0 = \alpha_1 = \alpha_2 = \alpha_3 = 1/2$. We know that this could be accomplished by shaping the electron wavepacket[36] such that $\psi_0 = \psi_1 = \psi_2 = \psi_3 = 1/2$. However, we can improve the robustness of our system to different kinds of noise by widening the range of occupied electron energy levels, which can be done using complex laser pulses composed of multiple harmonics of the driving laser[36]. This could improve the signal to noise ratio (SNR) if the electron energy spectrum is shaped to approximate a qudit state from equation (3). The challenge of the method of using laser pulses composed of multiple harmonics is the generation of this set of different laser frequencies with specific values of relative phases

and relative amplitudes. An alternative approach could rely on using multiple points of interaction with pre-determined distances between them[47].

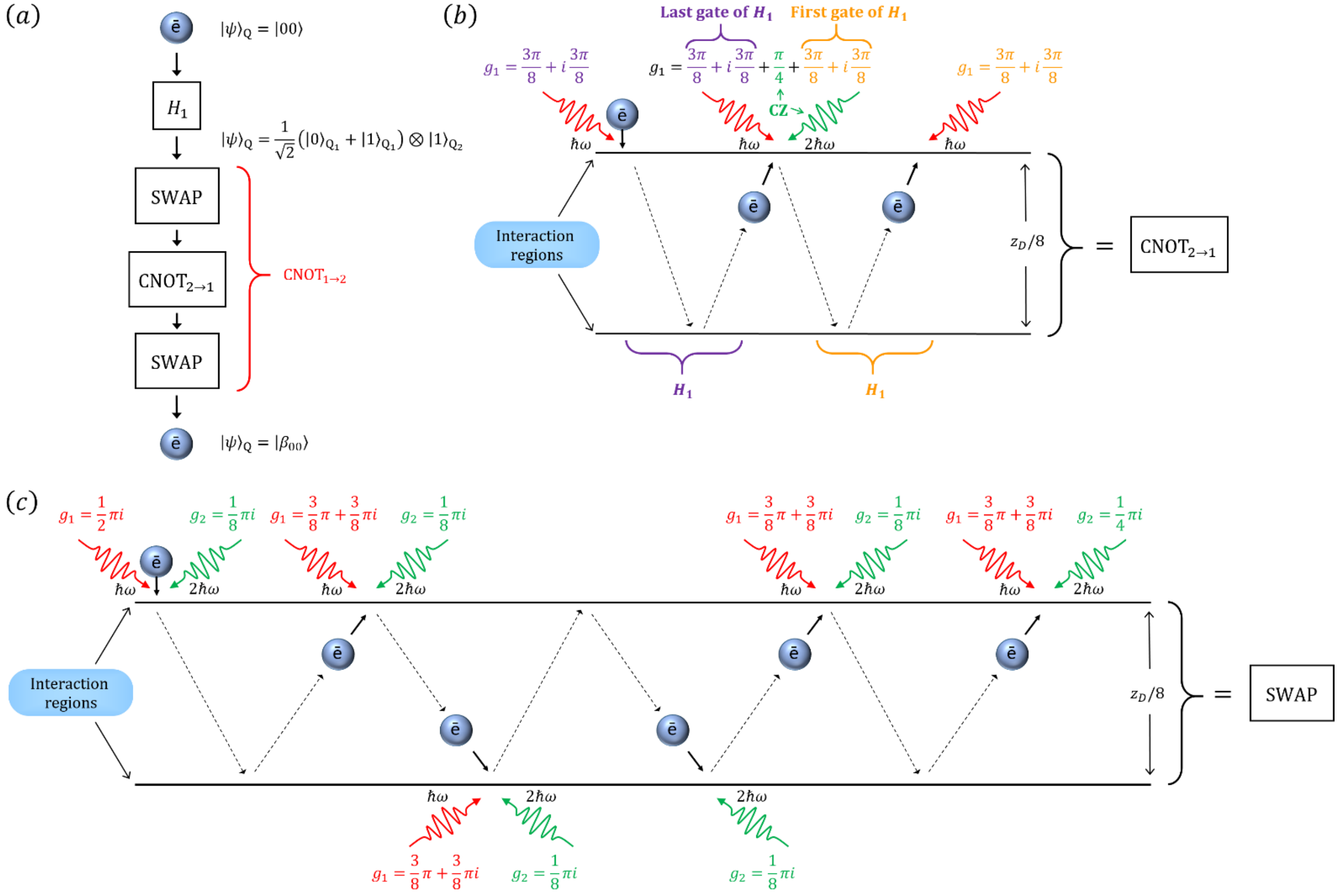

**Figure 3: Creating a free-electron Bell state $|\boldsymbol{\beta}_{00}\rangle_{\mathbf{Q}} \equiv \frac{(|00\rangle+|11\rangle)}{\sqrt{2}}$. (a)** An initial electron with the qubits state $|00\rangle_{\mathrm{Q}}$ undergoes a certain combination of PINEM and FSP opeartors and ends up in the maximally entangled state $|\beta_{00}\rangle_{\mathrm{Q}}$. **(b)** The implementaion of the $\mathrm{CNOT}_{2\to1}$ gate, using 3 PINEM gates seperated by free-space propagations of different distances. **(c)** The implementaion of the SWAP gate, using 6 PINEM gates seperated by free-space propagations of different distances. The interactions can be implemented by the electron moving in one direction, having multiple interaction points along its path, or by multi-pass electron microscopy with a pre-designed sequence of pulses at the interaction point(s).

## Results: extensions to an $n$-qubit scheme

So far, we have shown how to realize a 4-dimensional qudit (2 qubits subspaces). The natural question that arises is: **can an extension be made to the realization of a qudit of dimension $d = 2^n$, where $n$ is a natural number corresponding to the number of qubit subspaces?** To answer this question, we generalize equation (3) to define the qudit as:

$$|\psi\rangle_{\mathrm{Q}} \equiv \begin{pmatrix} \alpha_0 \\ \alpha_1 \\ \vdots \\ \alpha_{d-1} \end{pmatrix} = \frac{1}{\sqrt{d}} \begin{pmatrix} \sum_\ell \zeta_d^{0\cdot\ell} \cdot \psi_\ell \\ \sum_\ell {\zeta_d}^{1\cdot\ell} \cdot \psi_\ell \\ \vdots \\ \sum_\ell {\zeta_d}^{(d-1)\cdot\ell} \cdot \psi_\ell \end{pmatrix}, \tag{5}$$

where $\zeta_d = e^{-2\pi i/d}$ is a primitive $d$-th root of unity. We note a similarity to the $d$-dimensional Discrete Fourier Transform (DFT) matrix. For this choice of the qudit state, the PINEM matrix representation is diagonal. We denote by $g_j$ the coupling constant associated with the $j$-th harmonic $\omega_j = j \cdot \omega$. The PINEM matrix for $J$ laser harmonics $\{\omega, 2\omega, 3\omega, \dots, J\omega\}$ is

$$U_{PINEM}\left(\{g_j\}_{j=1}^{J}\right) = \begin{pmatrix} \lambda_0 & & & \\ & \lambda_1 & & \\ & & \ddots & \\ & & & \lambda_{d-1} \end{pmatrix}, \tag{6}$$

where the diagonal elements are

$$\lambda_k = \exp\left(2i \sum_{r=1}^{J} \frac{1}{r} |g_r| \sin\left(r \cdot \frac{2\pi k}{d} - \arg(g_r)\right)\right), \qquad k = 0,1,\dots,d-1. \tag{7}$$

The $d$-dimensional FSP gate is given as a $d$-dimensional DFT transformation on the diagonal form of the FSP, using the fundamental FSP distance $z = z_D/2d$:

$$U_{FSP}\left(z = \frac{z_D}{2d}\right) = U_{DFT}^{\dagger} \cdot \widetilde{U}_{FSP}\left(z = \frac{z_D}{2d}\right) \cdot U_{DFT}, \tag{8}$$

where $[U_{DFT}]_{j,k} = \frac{1}{\sqrt{d}}{\zeta_d}^{j \cdot k}$, and $A^{\dagger}$ is the conjugate transpose of $A$, and

$$\widetilde{U}_{FSP}\left(z = \frac{z_D}{2d}\right) = \begin{pmatrix} e^{-i\pi \cdot \frac{0^2}{d}} & & & & & \\ & e^{-i\pi \cdot \frac{1^2}{d}} & & & & \\ & & e^{-i\pi \cdot \frac{2^2}{d}} & & & \\ & & & \ddots & & \\ & & & & e^{-i\pi \cdot \frac{(d-2)^2}{d}} & \\ & & & & & e^{-i\pi \cdot \frac{(d-1)^2}{d}} \end{pmatrix}. \tag{9}$$

The remaining challenge is proving the universality of these operations for a general number of qubits, i.e., that the PINEM and FSP operations are sufficient to transform an initial state to any point in Hilbert space. Below, we present the steps necessary for such a proof, emphasizing what parts are still unproven – given as **conjectures 1-3**. These conjectures generalize the findings in lower dimensions and rely on the proven results summarized in the next section.

**Conjecture 1:** Given a $d$-dimensional qudit space, where $d = 2^n$, and the PINEM and FSP operators (equations (5-9)), there exists a set of PINEM coupling constants and a FSP propagation distance, such that these gates operate only on a portion of the qubit subspaces. The PINEM gate operates on the *rightmost* qubit subspaces, while the FSP operates on *leftmost* qubit subspaces. Formally:

$$U_{PINEM}^{(n\text{ qubits})}\left(\left\{g_j^{(k)}\right\}\right) = \mathbb{I}_2^{\otimes k} \otimes U_{PINEM}^{((n-k)\text{ qubits})}\left(\left\{g_j^{(k)}\right\}\right)$$

$$U_{FSP}^{(n\text{ qubits})}\left(z=\frac{z_D}{2\cdot 2^n/2^k}\right)=U_{FSP}^{((n-k)\text{ qubits})}\left(z=\frac{z_D}{2\cdot 2^n/2^k}\right)\otimes \mathbb{I}_2^{\otimes k}$$

where $k\in\{0,1,\dots,n-1\}$ and $\left\{g_j^{(k)}\right\}$ is a certain set of the PINEM coupling constants. Moreover, we conjecture that the $n$-qubit PINEM can operate on the rightmost qubit subspace when only the highest coupling constant is considered (see results (I) and (V)).

**Conjecture 2:** Given enough PINEM harmonics $\omega_j$, the PINEM operator is a universal phase gate. Mathematically, for each phase gate $U=\exp\left(i\ \mathrm{diag}(\varphi_0,\dots,\varphi_{d-1})\right)$ there exists a PINEM operation with frequencies $\{g_j\}$ such that $U=U_{PINEM}(\{g_j\})$ up to a global phase. Specifically, from combinatorial considerations, the set $\left\{g_1,\dots,g_{\left\lfloor\frac{d}{2}\right\rfloor}\right\}$ should be sufficient.

**Implication:** Together, **conjectures 1 and 2** imply universality on the leftmost two qubits.

**Conjecture 3:** SWAP gates can be implemented between each two (algebraically) nearest-neighbor qubits.

**Implication:** PINEM and FSP create a universal set of gates for $\mathrm{SU}(d)$.

An important question for future research concerns with the number of harmonics required to obtain a target state in the qudit Hilbert space. We found that a single laser harmonic is sufficient for implementing a single qubit[9], and two laser harmonics are required for creating any desirable state in a 2-qubit space. It is currently unknown how many laser harmonics are required for larger subspaces, e.g., that are equivalent to $n$ qubits. It could be scaling linearly or exponentially in the number of qubits (i.e., log of the dimension or proportional to the dimension).

## Summary of results proven for laser-driven free electron states

We summarize several mathematical results describing a decomposition of the PINEM and FSP gates to lower-dimension gates, i.e., operating just on a portion of the qubits subspaces. Full details are provided in the supplementary material (see Supplementary material section 8.6).

I. PINEM gate, 2 qubits → 1 qubit:

$$U_{PINEM}^{(2\ qubits)}(g_1 = 0, g_2) = \mathbb{I}_2 \otimes U_{PINEM}^{(1\ qubit)}(g_2)$$

$$U_{PINEM}^{(2\ qubits)}\left(g_1 = |g_1|e^{i\pi/4}, g_2 = 0\right) = U_{PINEM}^{(1\ qubit)}(g_2) \otimes \mathbb{I}_2$$

II. FSP gate, 2 qubits → 1 qubit:

$$U_{FSP}^{(2\text{ qubits})}\left(z = \frac{z_D}{4}\right) = \left[U_{FSP}^{(2\text{ qubits})}\left(z = \frac{z_D}{8}\right)\right]^2 = U_{FSP}^{(1\text{ qubit})}\left(z = \frac{z_D}{4}\right) \otimes \mathbb{I}_2$$

III. PINEM gate, 3 qubits → 2 qubits:

$$U_{PINEM}^{(3\text{ qubits})}(g_1 = 0, g_2, g_4) = \mathbb{I}_2 \otimes U_{PINEM}^{(2\text{ qubits})}(g_2, g_4)$$

IV. FSP gate, 3 qubits → 2 qubits:

$$U_{FSP}^{(3\text{ qubits})}\left(z = \frac{z_D}{8}\right) = \left[U_{FSP}^{(3\text{ qubits})}\left(z = \frac{z_D}{16}\right)\right]^2 = U_{FSP}^{(2\text{ qubits})}\left(z = \frac{z_D}{8}\right) \otimes \mathbb{I}_2$$

V. PINEM gate, 3 qubits → 1 qubit:

$$U_{PINEM}^{(3\text{ qubits})}(g_1 = 0, g_2 = 0, g_4) = \mathbb{I}_2 \otimes \mathbb{I}_2 \otimes U_{PINEM}^{(1\text{ qubit})}(g_4)$$

VI. FSP gate, 3 qubits → 1 qubit:

$$U_{FSP}^{(3\text{ qubits})}\left(z = \frac{z_D}{4}\right) = \left[U_{FSP}^{(3\text{ qubits})}\left(z = \frac{z_D}{16}\right)\right]^4 = U_{FSP}^{(1\text{ qubits})}\left(z = \frac{z_D}{4}\right) \otimes \mathbb{I}_2 \otimes \mathbb{I}_2$$

Note that in all the results presented here except result (I), the dimensionally-reduced PINEM (FSP) operates on the *rightmost (leftmost)* qubit subspace.

## Discussion: toward an experimental implementation of the concept

An experimental realization requires PINEM interactions, separated by one or more stretches of free-space propagation. These are the two important components that are needed to create the operators described in equations (1) and (2), from which a universal set of quantum gates can be realized. The magnitude of the coupling constants required to implement the universal set of gates is on the order of $|g_{1,2}| \sim \pi,$ which in the optical range is equivalent to electric field on the order of $10 - 100\ \mathrm{MV/m}$. Fields in this range are by now regularly used in UTEM experiments (e.g., [30,31,33,35,37,41]). The distances of the free-space propagation are in the range $0.1 - 1$ mm $(\sim z_D/8, z_D)$ These distances are regularly used in experiments of free-electrons interacting with lasers[32-34]. Quantum information processing with free electrons has recently been demonstrated experimentally for the specific case of a single free-electron qubit, utilizing interactions and free-space propagation[10]. This shows the feasibility of the schemes that we present in this work.

Envisioning experimentally feasible realizations, we take as a reference recent works[42,48] that demonstrated PINEM interactions covering up to ~1000 energy levels on a single electron in the UTEM. In an optimal scenario, this corresponds to a 1000-sized qudit or $\log_2 1000 \approx 10$ qubits on a single electron. We note that the analysis presented in this manuscript still holds when describing ~1000 electron energy levels, since electrons in UTEM setups have typical kinetic energies of 100-200 keV, which is 5 orders of magnitude larger than the typical photon energy $\hbar\omega$ (order of 1eV). Therefore, the lowest level is still more than 3 orders of magnitude higher than the energy separation $\hbar\omega$ (far above the absolute bottom of the electron energy ladder where its kinetic energy is zero).

To extract the electron state, the probability of each electron energy state in the ladder can be obtained using electron energy loss spectroscopy (EELS) that exists in many UTEM systems[30,35,49]. The phase of each energy component can be obtained using an additional PINEM interaction, which was already demonstrated in the UTEM[33]. We can then reconstruct the qudit state according to equation (3). Altogether, it becomes possible to implement full quantum state tomography of the electron quantum state. A different approach for the implementation of free-electron qubits has been recently demonstrated experimentally[10], and can also be generalized to contain higher dimensional quantum states.

Another promising direction for implementing electron-laser interactions in an especially scalable way is multi-pass electron microscopy[50,51], now being developed for other applications of quantum electron microscopy[52]. This scheme enables the electron to periodically revisit a single interaction point along its trajectory, enabling serial PINEM interactions alternating with FSP at the required fixed distances, as shown in figure 2 and figure 3.

## Discussion: computation with a qudit vs multiple qubits

A note should be made on the difference between realizing a single-qudit and realizing a multi-qubit scheme on a single electron. The similarly is expressed via the algebraic correspondence between a $d$-sized qudit and $\log_2 d$ qubits since they both represent the same $\mathrm{SU}(d)$ algebra lying in the same $d$-dimenisonal Hilbert space. The difference arises from a practical consideration related to the measurement: It is more complicated to measure one qubit out of multiple qubits defined on a single electron, since all qubits are carried on the same particle in the same physical degree of freedom. However, this is not a fundamental difference, because such a measurement is in principle possible. For example, measurement of a single-qubit subspace of the Hilbert space of the electron qudit can be done by designing an electron spectrometer that does not cause a complete collapse, but only projects the electron to the particular subspace, while keeping part of its coherence (e.g., using a barrier with multiple slits). The reality of such a possibility touches on a fundamental aspect of quantum mechanics – on the physics of collapse and mechanisms by which a wavefunction only partially collapses.

Another example where a higher dimensional Hilbert space can be partially measured without losing the quantum information is in the case of Gottesman-Kitaev-Preskill (GKP) states[22], in continuous-variable quantum information. There, a certain measurement can take part in error-correction without altering the (potentially unknown) encoded qubit state. We expect similar possibilities to be feasible using laser-driven free electrons, altering the theory to account for the different commutation relations in the free-electron synthetic Hilbert space.

The concept of having multiple qubits on the same free electron differs from other realizations of multiple qubits by having them encoded on the *same* physical system (single electron). The fundamental difference between having multiple qubits and having a larger qudit space is that the dimensions comprising each qubit are not *physically* separated, but only *algebraically* separated. Further enlarging the space of qudits can be considered using several electrons, each carrying a single qudit. Creating connectivity between different electrons was suggested in [53] by using efficient coupling to photonic cavities[41,42].

Most implementations of quantum information processing nowadays use multiple qubits; however, some implementations utilize fewer higher-dimensional qudits[11,12,54-57] for the same quantum operations[58,59]. In some examples in the literature, a single qudit could simulate quantum computations of numerous qubits[13,58,60-63]. For this, the qudit's dimension must be big enough to fully encode the information for the entire computation. Several papers in the literature show the feasibility of using a single large-sized qudit for implementing complex quantum algorithms. A $2^n$-dimensional qudit with a universal set of operations can simulate any unitary operation that acts within this $2^n$-dimensional Hilbert space. Nevertheless, due to the challenge of measuring a qubit subspace without collapsing the rest of space, it is not yet known whether all multiple-qubit computations could be implemented on a single electron.

## Discussion: comparing to other implementation schemes

Let us compare free electrons to bound-electron systems to motivate the consideration of free electrons as physical carriers of quantum information. Electrons operate in orders-of-magnitude higher energy scales and shorter time scales[30,35,49], yet the method for encoding information in them are remarkably different[9,33,35,36]. Compared to photonic systems, free electrons enable different interaction types and operations[25-26].

Another advantage of encoding quantum information on laser-driven free electrons is the *configurability* of the information encoding structure and its processing. By only changing of laser pulse properties, it is possible to switch between different gates. Looking forward, one can envision encoding an entire algorithm as a sequence of laser pulses.

It is instructive to also compare our free-electron-based qudit to mathematically related concepts that rely on infinite ladders of states. There has been extensive research on free-electron OAM manipulation[64-71], as well as photonic OAM[72], yielding high-quality generation and control of OAM states. These states have already been proposed to be utilized for encoding quantum information of the synthetic dimension of OAM[73,74]. Compared to our scheme, current schemes for OAM-quantum-information are similar in harnessing the infinite Hilbert space of states, but differ in the technical methods used for modulating OAM-carrying light beams, such as beam splitters and spatial light modulators. Despite the progress in OAM quantum information processing, an OAM-based framework was not yet shown to fulfill all the DiVincenzo criteria. It may be possible to learn from the schemes developed in our work and develop analogous schemes on the OAM synthetic dimension.

## Outlook

To conclude, this paper developed a novel idea for manipulating quantum information on a free electron by "Hilbert space engineering". This idea drew its motivation from a recent theoretical proposal[9] for encoding a qubit on free electrons using their interaction with ultrafast laser pulses. This proposal then formed the basis of an experiment which showed the formation of such free-electron qubits, reported in Ref. [10]. This demonstration relied on the two ingredients discussed also in this paper: laser–free-electron interaction and free-space propagation. While in [9] the qubit was mathematically defined using the infinite sum of odd and even probability amplitudes $f_l$, Ref. [10] took on a different approach of defining the qubit using only $f_{l=1}$ and $f_{l=-1}$.

Our research direction offers new applications for multi-pass electron microscopy, especially when driven by lasers like the UTEMs. Looking forward, having a fully controllable free-electron qudit may enable new analytical capabilities in electron microscopes. For example, probing the coherent state of quantum matter and its decoherence rates[25]. It is intriguing to explore what information could be extracted from a specimen by using a free-electron qudit as a probe for imaging or energy loss spectroscopy. Future research can also study changes of the free-electron qudit state by different elastic and inelastic interactions, e.g., with plasmons, excitons, and phonons[27].

## References

[1] Nielsen, M. A., & Chuang, I. Quantum computation and quantum information (2002).

[2]. Cirac, J. I., & Zoller, P. Quantum computations with cold trapped ions. *Physical review letters* **74**, 4091 (1995).

[3] Loss, D., & DiVincenzo, D. P. Quantum computation with quantum dots. *Physical Review A* **57**, 120 (1998).

[4] Knill, E., Laflamme, R., & Milburn, G. J. A scheme for efficient quantum computation with linear optics. *Nature* **409**, 46 (2001)

[5] Devoret, M. H., & Schoelkopf, R. J. Superconducting circuits for quantum information: an outlook. *Science*, **339**, 1169 (2013).

[6] Takui, T., Berliner, L., & Hanson, G. Electron spin resonance (ESR) based quantum computing. Springer New York (2016).

[7] P. W. Shor, Algorithms for quantum computation: discrete logarithms and factoring. *Proceedings 35th Annual Symposium on Foundations of Computer Science*, Santa Fe, NM, USA, pp. 124-134. (1994).

[8] Saffman, M., Walker, T. G., & Mølmer, K. Quantum information with Rydberg atoms. *Reviews of modern physics* **82**, 2313 (2010).

[9] Reinhardt, O., Mechel, C., Lynch, M. H., & Kaminer, I. Free-Electron Qubits, *Annalen der Physik* **533**, 2000254 (2020).

[10] Tsarev, M. V., Ryabov, A., & Baum, P. Free-electron qubits and maximum-contrast attosecond pulses via temporal Talbot revivals. *Physical Review Research* **3**, 043033 (2021).

[11] Wernsdorfer, W., & Ruben, M. Synthetic Hilbert Space Engineering of Molecular Qudits: Isotopologue Chemistry. *Advanced Materials* **31**, 1806687 (2019).

[12] Sawant, R., Blackmore, J. A., Gregory, P. D., Mur-Petit, J., Jaksch, D., Aldegunde, J., ... & Cornish, S. L. Ultracold polar molecules as qudits. *New Journal of Physics* **22**, 013027 (2020).

[13] Lu, H. H., Hu, Z., Alshaykh, M. S., Moore, A. J., Wang, Y., Imany, P., ... & Kais, S. Quantum Phase Estimation with Time-Frequency Qudits in a Single Photon. *Advanced Quantum Technologies* **3**, 1900074 (2020).

[14] J. Brendel, N. Gisin, W. Tittel, H. Zbinden. Pulsed energy-time entangled twin-photon source for quantum communication. *Physical Review Letters* **82**, 2594 (1999).

[15] I. Marcikic, H. De Riedmatten, W. Tittel, H. Zbinden, M. Legré, N. Gisin. Distribution of time-bin entangled qubits over 50 km of optical fiber. *Physical Review Letters* **93**, 180502 (2004).

[16] Babazadeh, A., Erhard, M., Wang, F., Malik, M., Nouroozi, R., Krenn, M., & Zeilinger, A. High-dimensional single-photon quantum gates: concepts and experiments. *Physical review letters* ***119***, 180510 (2017)

[17] Kok, P., Munro, W. J., Nemoto, K., Ralph, T. C., Dowling, J. P., & Milburn, G. J. Linear optical quantum computing with photonic qubits. *Reviews of Modern Physics* **79**, 135 (2007).

[18] O'brien, J. L. Optical quantum computing. *Science* **318**, 1567 (2007).

[19] O'brien, J. L., Furusawa, A., & Vučković, J. Photonic quantum technologies. *Nature Photonics* **3**, 687 (2009).

[20] Aspuru-Guzik, A., & Walther, P. Photonic quantum simulators. *Nature physics* **8**, 285 (2012).

[21] Lloyd, S., & Braunstein, S. L. Quantum computation over continuous variables. In Quantum Information with Continuous Variables. Springer, Dordrecht. (1999).

[22] Gottesman, D., Kitaev, A. & Preskill, J. Encoding a qubit in an oscillator. *Phys. Rev. A* **64**, 12310 (2001).

[23] Walschaers, M. Non-Gaussian Quantum States and Where to Find Them. *PRX Quantum* **2**, 30204 (2021).

[24] Arrazola, J. M. et al. Quantum circuits with many photons on a programmable nanophotonic chip. *Nature* **591**, 54 (2021).

[25] Ruimy, R., Gorlach, A., Mechel, C., Rivera, N., & Kaminer, I. Toward Atomic-Resolution Quantum Measurements with Coherently Shaped Free Electrons. *Physical Review Letters* **126**, 233403 (2021).

[26] Zhao, Z., Sun, X. Q., & Fan, S. Quantum entanglement and modulation enhancement of free-electron-bound-electron interaction. *Phys. Rev. Lett.* **126**, 233402 (2021).

[27] Mechel, C., Kurman, Y., Karnieli, A., Rivera, N., Arie, A., & Kaminer, I. Quantum correlations in electron microscopy. *Optica* **8**, 70 (2021).

[28] Kruit, P., Krielaart, M., van Staaden, Y., & Loginov, S. Creating contrast in electron microscopy using the quantum Zeno effect. EBSN 2019.

[29] Zewail, A. H. Four-dimensional electron microscopy. *Science* **328**, 187 (2010).

[30] Barwick, B., Flannigan, D. J., & Zewail, A. H. Photon-induced near-field electron microscopy. *Nature* **462,** 902 (2009).

[31] Echternkamp, K. E., Feist, A., Schäfer, S., & Ropers, C. Ramsey-type phase control of free-electron beams. *Nature Physics* **12**, 1000 (2016).

[32] Morimoto, Y., & Baum, P. Diffraction and microscopy with attosecond electron pulse trains. *Nature Physics* **14**, 252 (2018).

[33] Priebe, K. E., Rathje, C., Yalunin, S. V., Hohage, T., Feist, A., Schäfer, S., & Ropers, C. Attosecond electron pulse trains and quantum state reconstruction in ultrafast transmission electron microscopy. *Nature Photonics* **11**, 793 (2017).

[34] Kozák, M., Schönenberger, N., & Hommelhoff, P. Ponderomotive generation and detection of attosecond free-electron pulse trains. *Physical review letters* **120,** 103203 (2018).

[35] Feist, A., Echternkamp, K. E., Schauss, J., Yalunin, S. V., Schäfer, S., & Ropers, C. Quantum coherent optical phase modulation in an ultrafast transmission electron microscope. *Nature* **521,** 200 (2015).

[36] Reinhardt, O., & Kaminer, I. Theory of Shaping Electron Wavepackets with Light. *ACS Photonics* **7**, 2859 (2020).

[37] Vanacore, G. M., et al. Attosecond coherent control of free-electron wave functions using semi-infinite light fields. *Nature communications* **9**, 2694 (2018).

[38] García de Abajo, F. J., Asenjo-Garcia, A., & Kociak, M. Multiphoton absorption and emission by interaction of swift electrons with evanescent light fields. *Nano letters* **10,** 1859 (2010).

[39] Park, S. T., Lin, M., & Zewail, A. H. Photon-induced near-field electron microscopy (PINEM): theoretical and experimental. *New Journal of Physics* **12**, 123028 (2010).

[40] García de Abajo F. J., Barwick, B., & Carbone, F. Electron diffraction by plasmon waves. *Physical Review B* **94**, 041404 (2016).

[41] Wang, K., Dahan, R., Shentcis, M., Kauffmann, Y., Hayun, A. B., Reinhardt, O., ... & Kaminer, I. Coherent interaction between free electrons and a photonic cavity. *Nature* **582**, 50 (2020).

[42] Kfir, O. et al. Controlling free electrons with optical whispering-gallery modes. *Nature* **582**, 46 (2020)

[43] Jones, J. A., & Mosca, M. Implementation of a quantum algorithm on a nuclear magnetic resonance quantum computer. *The Journal of chemical physics* **109**, 1648 (1998).

[44] Vatan, Farrokh, and Colin Williams. Optimal quantum circuits for general two-qubit gates. *Physical Review A* **69,** 032315 (2004).

[45] Feist, A., et al. Ultrafast transmission electron microscopy using a laser-driven field emitter: Femtosecond resolution with a high coherence electron beam. *Ultramicroscopy* **176**, 63 (2017).

[46] Losquin, A., & Lummen, T. T. Electron microscopy methods for space-, energy-, and time-resolved plasmonics. *Frontiers of Physics* **12**, 127301 (2017).

[47] Yalunin, S. V., Feist, A., & Ropers, C. Tailored high-contrast attosecond electron pulses for coherent excitation and scattering. *Physical Review Research* **3**, 032036 (2021).

[48] Dahan, R., et al. Resonant phase-matching between a light wave and a free-electron wavefunction. *Nature Physics* **16**, 1123 (2020).

[49] Piazza, L. Lummen, T., Quinonez, E., Murooka, Y., Reed, B. W., Barwick, B., & Carbone, F. Simultaneous observation of the quantization and the interference pattern of a plasmonic near-field. *Nature communications* **6,** 6407 (2015).

[50] Kruit, P., et al. Designs for a quantum electron microscope. *Ultramicroscopy* **164**, 31 (2016).

[51] Juffmann, T., Koppell, S. A., Klopfer, B. B., Ophus, C., Glaeser, R. M., & Kasevich, M. A. Multi-pass transmission electron microscopy. *Scientific reports* **7**, 1699 (2017).

[52] Okamoto, H., Latychevskaia, T., & Fink, H. W. A quantum mechanical scheme to reduce radiation damage in electron microscopy. *Applied physics letters* **88**, 164103 (2006).

[53] Kfir, O. Entanglements of electrons and cavity photons in the strong-coupling regime. *Physical review letters* **123**, 103602 (2019).

[54] Imany, P., Jaramillo-Villegas, J. A., Alshaykh, M. S., Lukens, J. M., Odele, O. D., Moore, A. J., ... & Weiner, A. M. High-dimensional optical quantum logic in large operational spaces. *npj Quantum Information* **5**, 1 (2019).

[55] Lu, H. H., Weiner, A. M., Lougovski, P., & Lukens, J. M. Quantum information processing with frequency-comb qudits. *IEEE Photonics Technology Letters* **31**, 1858 (2019).

[56] Kues, M., Reimer, C., Roztocki, P., Cortés, L. R., Sciara, S., Wetzel, B., ... & Moss, D. J. On-chip generation of high-dimensional entangled quantum states and their coherent control. *Nature* **546**, 622 (2017).

[57] Low, P. J., White, B. M., Cox, A. A., Day, M. L., & Senko, C. Practical trapped-ion protocols for universal qudit-based quantum computing. *Physical Review Research* **2**, 033128 (2020).

[58] Bremner, M. J., Bacon, D., & Nielsen, M. A. Simulating Hamiltonian dynamics using many-qudit Hamiltonians and local unitary control. *Physical Review A* **71** 052312 (2005).

[59] Rungta, P., Munro, W. J., Nemoto, K., Deuar, P., Milburn, G. J., & Caves, C. M. Qudit entanglement. In *Directions in Quantum Optics* (pp. 149-164). Springer, Berlin, Heidelberg. (2001).

[60] Pirandola, S., Mancini, S., Braunstein, S. L., & Vitali, D. Minimal qudit code for a qubit in the phase-damping channel. *Physical Review A* **77**, 032309. (2008).

[61] Marques, B., Matoso, A. A., Pimenta, W. M., Gutiérrez-Esparza, A. J., Santos, M. F., & Pádua, S. Experimental simulation of decoherence in photonics qudits. *Scientific reports* **5**, 16049 (2015).

[62] Gedik, Z., Silva, I. A., Çakmak, B., Karpat, G., Vidoto, E. L. G., Soares-Pinto, D. D. O., ... & Fanchini, F. F. Computational speed-up with a single qudit. *Scientific reports* **5**, 14671 (2015).

[63] Kiktenko, E. O., Fedorov, A. K., Strakhov, A. A., & Man'Ko, V. I. Single qudit realization of the Deutsch algorithm using superconducting many-level quantum circuits. *Physics Letters A* **379**, 1409 (2015).

[64] Bliokh, K. Y., Bliokh, Y. P., Savel'Ev, S., & Nori, F. Semiclassical dynamics of electron wave packet states with phase vortices. *Physical Review Letters* **99**, 190404 (2007).

[65] Cai, W., Reinhardt, O., Kaminer, I., & de Abajo, F. J. G. Efficient orbital angular momentum transfer between plasmons and free electrons. *Physical Review B* **98**, 045424 (2018).

[66] Vanacore, G. M., et al. Ultrafast generation and control of an electron vortex beam via chiral plasmonic near fields. *Nature materials* **18**, 573 (2019).

[67] Verbeeck, J., Tian, H., & Schattschneider, P. Production and application of electron vortex beams. *Nature* **467,** 301 (2010).

[68] Uchida, M., & Tonomura, A. Generation of electron beams carrying orbital angular momentum. *Nature* **464,** 737 (2010).

[69] McMorran, B. J., Agrawal, A., Anderson, I. M., Herzing, A. A., Lezec, H. J., McClelland, J. J., & Unguris, J. Electron vortex beams with high quanta of orbital angular momentum. *Science* **331,** 192 (2011).

[70] Larocque, H., Kaminer, I., Grillo, V., Leuchs, G., Padgett, M. J., Boyd, R. W., Segev M. & Karimi, E. 'Twisted' electrons. *Contemporary Physics* **59**, 126 (2018).

[71] Wang, J., Yang, J., Fazal, I. M., Ahmed, N., Yan, Y., Huang, H., Ren, Y., Yue, Y., Dolinar, S., Tur, M., Willner, A. E. Terabit free-space data transmission employing orbital angular momentum multiplexing. *Nature photonics* **6**, 488 (2012).

[72] Allen, L., Beijersbergen, M. W., Spreeuw, R. J. C., & Woerdman, J. P. Orbital angular momentum of light and the transformation of Laguerre-Gaussian laser modes. *Physical review A* **45**, 8185 (1992).

[73] Gibson, G., Courtial, J., Padgett, M. J., Vasnetsov, M., Pas'ko, V., Barnett, S. M., & Franke-Arnold, S. Free-space information transfer using light beams carrying orbital angular momentum. *Optics express* **12**, 5448 (2004).

[74] Perumangatt, C., Lal, N., Anwar, A., Reddy, S. G., & Singh, R. P. Quantum information with even and odd states of orbital angular momentum of light. *Physics Letters A* **22**, 1858 (2017).

## AUTHOR CONTRIBUTIONS

G. B., and O. R. conceived the idea and cowrote most of the paper. G. B., O. R., C. M., O. L., and I. K. all participated in discussions that developed the theory and shaped the project.

## COMPETING INTERESTS

The authors declare no competing interests.

## DATA AVAILABILITY

The data that support the findings of this study are available from the corresponding author upon a reasonable request.